\documentclass[paper]{raa} 

\usepackage{graphicx,times} 
\usepackage{natbib}
\usepackage{amssymb,amsmath}
\bibpunct{(}{)}{;}{a}{}{,}

\usepackage[a4paper=true,dvipdfm=true,pagebackref=true]{hyperref}
\hypersetup{colorlinks = true, linkcolor = green, anchorcolor = red, citecolor = blue, filecolor = red, pagecolor = red, urlcolor = red}

\begin{document}

\title{Photometric light curve solutions of three ultra-short period eclipsing binaries
}

\volnopage{Vol.0 (20xx) No.0, 000--000} 
\setcounter{page}{1} 

 \author{F. Acerbi\inst{1}, R. Michel\inst{2}, C. Barani\inst{1}, M. Martignoni\inst{1}, L. Fox-Machado\inst{2}}

\institute{Stazione Astronomica Betelgeuse, Magnago Italy; {\it acerbifr@tin.it}\\
\and
 Instituto de Astronom\'{i}a, Universidad Nacional Aut\'onoma de M\'exico\\
\vs\no
{\small Received~~20xx month day; accepted~~20xx~~month day}
}

\abstract{ We present the results of our study of the eclipsing binary systems CSS J112237.1+395219, LINEAR 1286561 and LINEAR 2602707 based on new CCD $B$, $V$, $R_c$ and $I_c$ complete light curves. The ultra-short period nature of the stars \citep{Drake2014} is confirmed and the system's periods are revised. The light curves were modelled using the 2005 version of the Wilson-Devinney code.
When necessary, cool spots on the surface of the primary component were introduced to account for asymmetries in the light curves. As a result, we found that CSS J112237.1+395219 is a W UMa type contact binary system belonging to W subclass with a mass ratio of $q=1.61$ and a shallow degree of contact of 14.8\% where the primary component is hotter than the secondary one by $500K$. LINEAR 1286561 and LINEAR 2602707 are detached binary systems with mass ratios $q=3.467$ and $q=0.987$ respectively. These detached systems are low-mass M-type eclipsing binaries of similar temperatures.
The marginal contact, the fill-out factor and the temperature difference between components of CSS J112237.1+395219 suggest that this system may be at a key evolutionary state predicted by the Thermal Relaxation Oscillation theory (TRO).
From the estimated absolute parameters we conclude that our systems share common properties with others ultra-short period binaries.
\keywords{techniques: photometric --- stars: variables: Contact Binary --- stars: individual: CSS J112237.1+395219, LINEAR 1286561 and LINEAR 2602707}
}

\authorrunning{Acerbi et al.} 
\titlerunning{Light curve solutions of three ultra-short eclipsing binaries} 

\maketitle

\section{Introduction} 
\label{sect:intro}

The term ultra-short period binaries (USPBs) refers to binaries with orbital periods shorter than $\sim0.22d$ (Ruci\'nski  1992; 2007). Although several theories to explain the observed short-period cut-off have been proposed  \cite{Stepien2011}; \cite{Jiang2012}, the explanation of this abrupt short period limit is still an open question. Since Ruci\'nski works, several hundred  USPBs have been discovered, being the first one BW3V38 -- a detached system \cite{Udalski1995} while the list continues growing (e.g. \cite{Norton2011}, \cite{Prsa2011}, \cite{Lohr2013}, \cite{Drake2014}, \cite{Soszynski2015} and \cite{Li2017}). The short-period contact systems, especially those on the short-period end, are of great interest for the study of the structure and evolution of eclipsing binaries. These systems are expected to be composed of two K or later-type components, according to the period-color relation for contact binaries \citep{Zhu2015}. CSS J112237.1+395219 (hereinafter J112237), LINEAR 1286561  (hereinafter L1286561) and LINEAR 2602707  (hereinafter L2602707) were reported as USPBs systems with periods below 0.2 days by \cite{Drake2014}. The spectral type of the systems were obtained matching all the USPBs of their list with objects having spectra within the SDSS Data Release 10 \citep{Ahn2014}. Table \ref{table:list} lists the coordinates, periods and spectral types of our targets stars.

 J112237, being of K-type, is an important object for explaining the period cut-off phenomenon \citep{Liu2014a}. At present, however, only a few  binaries with periods shorter than 0.25 days have been studied in details. This makes our study on this \(<\)0.2 days period K-type system interesting. 
The other two systems are of M spectral type. Since short period M dwarf binaries are relatively faint objects, they are difficult to detect \citep{Davenport2013}. Therefore, any new detection of such a system constitute a valuable contribution to the understanding of eclipsing binary formation.

In this paper multicolor light curves are analyzed simultaneously using the 2003 version of the Wilson-Devinney Code, revision of the October 2005 (\cite{Wilson1971}; \cite{Wilson1990}; \cite{Wilson1994}; \cite{Wilson2004}. Asymmetries in the light curves of short-period binaries have been commonly reported and are attributed to spot activity on the stellar photospheres which can be modelled very well by hot or cool spots on the components of the systems.

\begin{table*}
\scriptsize
\begin{center}
\caption{List of targets}
\label{table:list}
\begin{tabular}{lcccc}
\hline\hline
Object & RA (2000) & DEC (2000) &  $P(d)$ & Spec. Type \\
\hline\noalign{\smallskip}
J112237  & 11:22:37.06 & +39:52:19.9 & 0.184749 & $K2V$ \\ 
L1286561 & 11:22:43.44 & +37:21:30.2 & 0.168744 & $M4.5V$ \\
L2602707 & 11:55:33.44 & +35:44:39.3 & 0.199725 & $M2V$ \\
\noalign{\smallskip}\hline
\end{tabular}
\end{center}
\end{table*}

\section{CCD Photometric observations and data reduction}
\label{sect:Obs}

Observations were carried out at the San Pedro Martir Observatory either at the 2m telescope fitted with a filter-wheel and the \textit{Marconi 4} CCD detector (a deep depletion e2v CCD42-40 chip with gain of 2.30 e$^-$/ADU and readout noise of 5.20 e$^-$) giving a field of view of $6^{\prime}\times6^{\prime}$ and at the 0.84-m telescope with another filter-wheel and the \textit{Spectral Instruments 1} CCD detector (a deep depletion e2v CCD42-40 chip with gain of 1.39 e$^-$/ADU and readout noise of 3.54 e$^-$) giving a $7.6^{\prime}\times7.6^{\prime}$ field of view. Binning 2$\times$2 was used during all the observations. Alternated exposures of the targets were taken with different Johnson-Cousins filters. The details of the observations are shown in Table \ref{table:obs}. Flat field and bias frames were also obtained during all the observing runs.

\begin{figure*}
\center
\includegraphics[width=12.0cm,height=4.0cm]{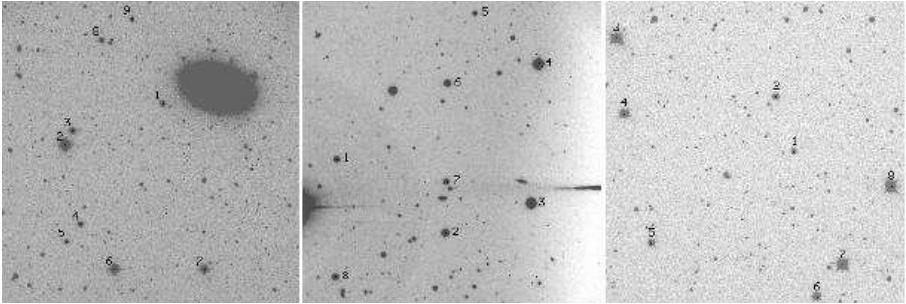}
\caption{Observed fields. J112237 the left, L1286561 in the middle and L2602707 to the right. The variable stars are marked with number 1.}
\label{fig:fields}
\end{figure*}

\begin{table*}
\scriptsize
\begin{center}
\caption{Log of photometric observations}
\label{table:obs}
\begin{tabular}{cccccccc}
\hline\hline
Target   &  UT Date    & Telescope & Obs.  & Filters  &  Exp. Times  &  Number of    & Mean errors \\
         &             &           & Hours &          &     [s]      &  Integrations &   [mag]     \\
\hline
J112237  & 2018 Apr  8 & 2m    &  4.9 & $B, V, R_c, I_c$ & 40, 20, 15, 15   & 105, 104, 106, 105 & 0.010, 0.009, 0.008, 0.011 \\
         \hline
L1286561 & 2018 Apr  6 & 2m    &  4.3 & $V, R_c, I_c$    & 30, 15, 10       & 130, 132, 131      & 0.012,  0.010, 0.009 \\
\hline
L2602707 & 2016 Mar 15 & 0.84m & 10.2 & $V, R_c, I_c$    & 120, 60, 30      & 122, 121, 110      & 0.009 , 0.009, 0.007 \\
         & 2018 Jan 16 & 0.84m &  2.0 & $V, R_c, I_c$    & 120, 60, 30      &  25,  24,  24      & 0.008 , 0.007, 0.005 \\
         & 2018 Apr  7 & 2m    &  0.7 & $V, R_c, I_c$    &  30, 20, 15      &  20,  19,  18      & 0.008 , 0.007, 0.008 \\
        \hline
\end{tabular}
\end{center}
\end{table*}

All images were processed using IRAF\footnote{IRAF is distributed by the National Optical Observatories (NOAO), operated by the Association of Universities for Research in Astronomy, Inc., under cooperative agreement with the National Science Foundation.} routines. Images were bias subtracted and flat field corrected before the instrumental magnitudes of the variables and some field stars were computed with the standard aperture photometry method. In order to choose good comparison stars (with similar colors to the variables), the fields were calibrated in the $UBV(RI)_c$ system (during very photometric nights) using standard stars from the Landolt catalog. The obtained magnitudes of the comparison stars are given in Table \ref{table:refs}. All or any part of the data is available upon request.

\begin{table*}
\scriptsize
\begin{center}
  \caption{$UBVR_cI_c$ magnitudes of the reference stars}
  \label{table:refs}
\begin{tabular}{ccrrccclllll}
\hline\hline
Field    &      Name         & RA (deg)   & DEC (deg)  &   $U$     &   $B$      &   $V$      &  $R_c$     & $I_c$       \\
\hline
J112237  & 2MASSJ11224702+3951334  & 170.696042 & +39.859296 & 15.531$\pm$0.010 & 15.714$\pm$0.016 & 15.403$\pm$0.022 & 14.754$\pm$0.002 & 14.306$\pm$0.030 \\
L1286561 & 2MASSJ11223284+3720037  & 170.636954 & +37.334328 & 19.273$\pm$0.023 & 18.327$\pm$0.007 & 16.710$\pm$0.003 & 15.650$\pm$0.002 & 14.400$\pm$0.003 \\
L2602707 & 2MASSJ11553585+3546032  & 178.899350 & +35.767600 & 17.657$\pm$0.023 & 16.773$\pm$0.010 & 15.713$\pm$0.007 & 15.103$\pm$0.006 & 14.542$\pm$0.005\\
\hline
\end{tabular}
\end{center}
\end{table*}

The new times of minima for the three systems, presented in Table~\ref{minima}, are all heliocentric and determined with the polynomial fit method. 

\begin{table}
\bc
\begin{minipage}[]{100mm}
\caption[]{CCD times of minima for the three systems.\label{minima}}\end{minipage}
\setlength{\tabcolsep}{1pt}
\small
 \begin{tabular}{ccccccccc}
\hline\hline
& Filters & HJD & Epoch(1) & O-C(1) & Epoch(2) & O-C(2) & Error & Source \\
  \hline\noalign{\smallskip}
J112237  & BVRI & 2458213.6800 &    -0.5 & -0.0027 &    -0.5 &  0.0000 & 0.0009 & This paper \\
         & BVRI & 2458213.7712 &     0.0 & -0.0039 &     0.0 &  0.0000 & 0.0013 & This paper \\
\hline
L1286561 &      & 2453711.9610 &     0.0 &  0.0000 &     0.0 &  0.0000 &        & VSX \\
         & VRI  & 2458214.6438 & 26683.5 &  0.0009 & 26683.5 & -0.0001 & 0.0009 & This paper \\
         & VRI  & 2458214.7284 & 26684.0 &  0.0011 & 26684.0 &  0.0001 & 0.0008 & This paper \\
\hline
L2602707 & VRI  & 2457462.6623 &    -0.5 &  0.0001 &    -0.5 &  0.0001 & 0.0010 & This paper \\
         & VRI  & 2457462.7621 &     0.0 &  0.0000 &     0.0 &  0.0000 & 0.0008 & This paper \\
         & VRI  & 2457462.8618 &     0.5 & -0.0002 &     0.5 & -0.0002 & 0.0007 & This paper \\
         & VRI  & 2457462.9617 &     1.0 & -0.0001 &     1.0 & -0.0001 & 0.0006 & This paper \\
         & VRI  & 2458135.0393 &  3366.0 &  0.0063 &  3366.0 &  0.0010 & 0.0006 & This paper \\
         & VRI  & 2458215.7266 &  3770.0 &  0.0050 &  3770.0 & -0.0009 & 0.0007 & This paper\\
  \noalign{\smallskip}\hline
\end{tabular}
\ec
\end{table}

These new data allowed us to refine the ephemeris of the systems as follows:

\begin{equation}
J112237, HJD (Min I) = 2458213.7712(5)+0d.1823999(4)\textit{E}.
\end{equation}
\begin{equation}
L1286561, HJD (Min I) = 2453711.9610(2) + 0d.1687441(1)\textit{E}.
\end{equation}
\begin{equation}
L2602707, HJD (Min I) = 2457462.7621(3) + 0d.1797256(2)\textit{E}.
\end{equation}

\section{Modelling the light curves}
\label{sect:Modelling}
The analyses of the observed light curves of the systems were carried out using the 2003 version (October 2005 revision) of the Wilson-Devinney code. To determine the mean surface temperature of star 1, we used the spectral classes provided by \cite{Drake2014} and, according to the  MK spectral types given in Allen’s Astrophysical Quantities \citep{Cox2000}, we assigned to the primary stars an effective temperature of $T_{eff} = 4830K$, $3120K$ and $3520K$ for J112237, L1286561 and L2602707 respectively and used them as primary temperatures in the light curve analyses.

In each case, the $q$\(-\)search method was performed to find the best initial value of $q$. We assumed gravity darkening and bolometric albedo exponents appropriate for the convective envelopes $(T_{eff} < 7500K)$. Limb-darkening coefficients of the components were interpolated from the square root law of the \cite{Claret2011} tables. 

 The shape of the light curve of J112237 resembles that of typical W UMa-type binary stars. Therefore, for this star we started the W-D analysis directly in Mode 3 - overcontact configuration.  For the other two systems, the calculations were started in Mode 2 - detached configuration. The different computation modes offered by the W-D code can be found in \cite{Wilson2004}.

The observed light curves of J112237 and L1286561 are asymmetric and show unequal quadrature heights, with maximum I being brighter than maximum II. A spotted model was introduced to account for the asymmetry in the observed light curves, known as O'Connell effect \citep{O'Connell1951}, which is  present in several eclipsing binary light curves and can be  explained by spot activity in the component stars \citep{Zhai1988} and are of the same nature as solar magnetic spots \citep{Mullan1975}. 

\begin{figure*}
\center
\includegraphics[width=18.0cm,height=6.0cm]{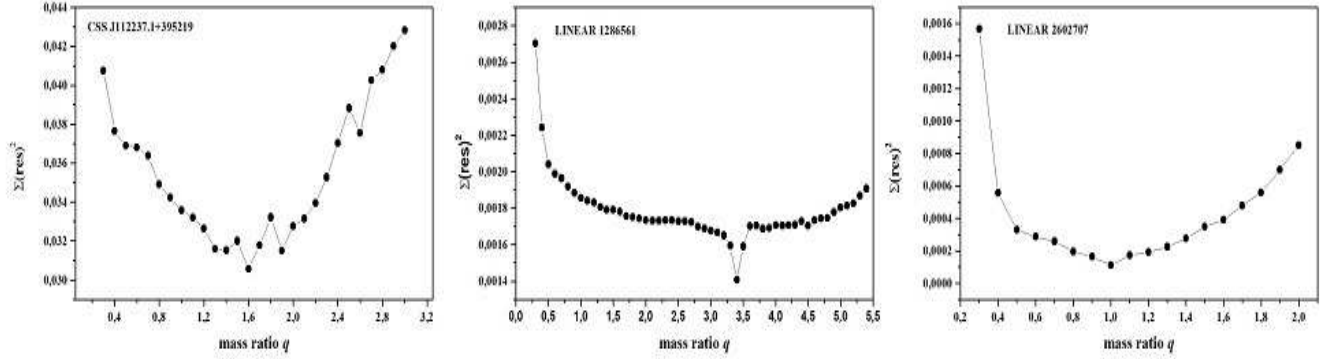}
\caption{The relation $\Sigma(res)^2$ versus mass ratio $q$ in Mode 3 for J112237 and in Mode 2 for the two LINEAR systems in the WD code.}
\label{fig:relation}
\end{figure*}

A sufficient number of runs of the DC program was made, until the corrections to the parameters became smaller than their probable errors. The corresponding relation is plotted in Figure~\ref{fig:relation}. 

\begin{table}
\bc
\begin{minipage}[]{100mm}
\caption[]{Light curve solutions for J112237 and the two LINEAR systems. Assumed parameters are marked with *.\label{pars}}\end{minipage}
\setlength{\tabcolsep}{1pt}
\small
 \begin{tabular}{cccccccc}
\hline\hline
Parameter  && J112237 && L1286561 && L2602707 \\
  \hline\noalign{\smallskip}
$i$ && 67.425$\pm$0.329 && 72.996$\pm$0.421 &  & 79.920$\pm$0.114 \\
$T_1(K)$ && 4830* && 3120* &  & 3520* \\
$T_2(K)$ && 4321$\pm$15 && 2989$\pm$20 &  & 3465$\pm$6 \\
$\Omega_1$ && 4.612$\pm$0.056 && 7.791$\pm$0.041 &  & 4.092$\pm$0.038 \\
$\Omega_2$ && 4.612$\pm$0.056 && 7.794$\pm$0.050 &  & 4.087$\pm$0.040 \\
$q= m_2/m_1$ && 1.616$\pm$0.035 && 3.468$\pm$0.065 &  & 0.988$\pm$ 0.021 \\
$A_1=A_2$ && 0.5* && 0.5* &  & 0.5* \\
$g_1=g_2$ && 0.32* && 0.32* &  & 0.32* \\
$L_{1B}$ && 0.533$\pm$0.004 &&  &  &\\
$L_{1V}$ && 0.510$\pm$0.004 && 0.260$\pm$0.005 &  & 0.513$\pm$0.009 \\
$L_{1Rc}$ && 0.476$\pm$0.005 && 0.253$\pm$0.007 &  & 0.509$\pm$0.008 \\
$L_{1Ic}$ && 0.455$\pm$0.004 && 0.247$\pm$0.007 &  & 0.508$\pm$0.008 \\
$L_{2B}$ && 0.371$\pm$0.003 &&  &  &\\
$L_{2V}$ && 0.396$\pm$0.002 && 0.691$\pm$0.004 &  & 0.451$\pm$0.008 \\
$L_{2Rc}$ && 0.442$\pm$0.004 && 0.699$\pm$0.005 &  & 0.455$\pm$0.008 \\
$L_{2Ic}$ && 0.470$\pm$0.003 && 0.728$\pm$0.003 &  & 0.462$\pm$0.008 \\
$f$ && 0.148$\pm$0.007 && -0.06$\pm$0.003 &  & -0.083$\pm$0.005  \\
$X_{1B}$ && 0.987* &&  &  &  & \\
$X_{1V}$ && 0.681* && 0.359* &  & 1.083* \\
$X_{1Rc}$ && 0.389* && 0.205* &  & 0.801* \\
$X_{1Ic}$ && 0.232* && -0.017* &  & 0.444* \\
$X_{2B}$ &&   &&  &  &\\
$X_{2V}$ &&   &  & 0.660* &  & -0.269* \\
$X_{2Rc}$ &&   &  & 0.785* &  & -0.003* \\
$X_{2Ic}$ &&   &  & 1.002* &  & 0.403* \\
$l_3$ && 0 &  & 0&  & 0\\
$r_1$ (pole) && 0.322$\pm$0.002 && 0.228$\pm$0.002 &  & 0.317$\pm$ 0.004 \\
$r_1$ (side) && 0.338$\pm$0.003 && 0.234$\pm$0.002 &  & 0.328$\pm$ 0.005 \\
$r_1$ (back) && 0.374$\pm$0.004 && 0.255$\pm$0.003 &  & 0.345$\pm$ 0.006 \\
$r_2$ (pole) && 0.403$\pm$0.007 && 0.428$\pm$0.002 &  & 0.315$\pm$ 0.004 \\
$r_2$ (side) && 0.428$\pm$0.009 && 0.452$\pm$0.003 &  & 0.326$\pm$ 0.004 \\
$r_2$ (back) && 0.460$\pm$0.014 && 0.469$\pm$0.003 &  & 0.343$\pm$ 0.007 \\
lat. spot && 60.34$\pm$2.1 && 85.07$\pm$1.9 & 95.27$\pm$2.2 &  &  \\
long. spot && 121.33$\pm$3.4 && 20.56$\pm$2.1 & 321$\pm$2.6 &  &  \\
radius && 25.3$\pm$0.86 && 30.33$\pm$0.73 & 21.13$\pm$0.90 &  &  \\
Temp .fact. && 0.925$\pm$0.021 && 0.940$\pm$0.036 & 0.851$\pm$0.045 &  &  \\
Component && 1 &  & 1   & 1 &  &\\
Sum(res)$^2$ && 0.0301 && 0.00138 &  & 0.00010 &  \\
  \noalign{\smallskip}\hline
\end{tabular}
\ec
\end{table}

\begin{figure*}
\center
\includegraphics[width=18.0cm,height=6.0cm]{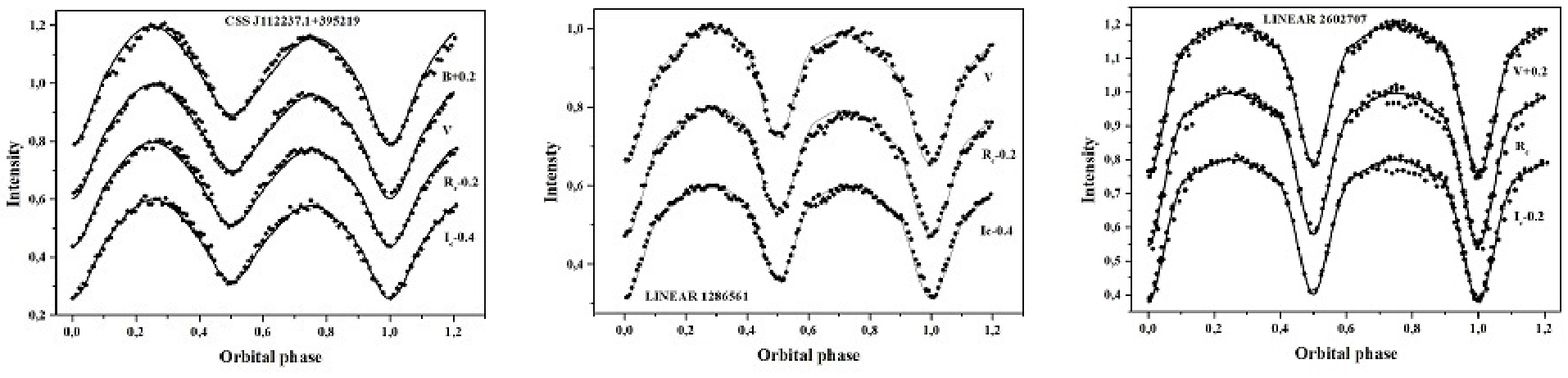}
\caption{CCD light curves of J112237 and the two LINEAR systems. Points are the original observations and lines the theoretical fit with the spot contribution, where due}
\label{fig:curves}
\end{figure*}

Starting with the preliminary solutions for the values of $q$ found, we performed a more detailed analysis with $q$ being treated as an additional free parameter. Parameters of the accepted solutions are listed in Table~\ref{pars}, while Figure~\ref{fig:curves} shows the best fits for the model parameter from WD code and the observed light curves of the systems. In our final solutions, we also found that the contribution of a third light is negligible.

It should be noted that the errors of the parameters given in this paper are the formal errors from the WD code and are known to be unrealistically small \citep{Maceroni1997}. For a discussion see \cite{Barani2017}.

\section{Estimate of the absolute elements}

Due to the lack of radial velocity (RV) solutions, we used empirical relations to determine the absolute parameters of the binary systems. \cite{Dimitrov2015} gave a period - semi-major axis ($P-a$) relation on the basis of 14 binary stars having $P<0.27d$ which had both RV and photometric solutions, which is approximated by a parabola:
\begin{equation}
a = -1.154 + 14.633 \x\ P -10.319 \x\ P^2.
\end{equation}
where\textit{ P} is in days and \textit{a }is in solar radii.

Following the above relation, we determined a semi-major axis $a(R_{\odot})$ of our three systems.

The $(P,a)$ relation (4) corresponds to the following relation "period-mass" for short-period binaries:
\begin{equation}
M= 0.0134 / P^2 \x\ (-1.154 + 14.633 \x\ P -10.319 \x\ P^2)^3.
\end{equation}
Where $M$ is the total mass of the binary.

The mean fractional radii of the components were obtained with the formula:

\begin{equation}
r_{1,2mean} =(r_{pole}+r_{side}+r_{back})^{1/3}.
\end{equation}

Using the semi-major axis, we can calculate the radii of the binary components as

\begin{equation}
R_{1,2} = a r_{1,2mean}.
\end{equation}

Considering a solar temperature of $T_{\odot}=5780K$, absolute parameters of bolometric magnitudes and luminosities can be calculated using the equations:
\begin{equation}
Mbol_{1,2} = 4.77 - 5log(R_{1,2}/R_{\odot}) - 10log(T_{1,2}/T_{\odot}).
\end{equation}

\begin{equation}
L_{1,2} = (R_{1,2} /R_{\odot})^2 \x\ (T_{1,2} / T_{\odot})^4.
\end{equation}

The mean densities of the binary components were derived from the following equation given by \cite{Mochnacki1981}:
\begin{equation}
\rho_1 = 0.0189/r_{1,mean}^3 P^2 (1+q).
\end{equation}

\begin{equation}
\rho_2 = 0.0189 q/r_{2mean}^3 P^2 (1+q).
\end{equation}

The mass of the primary component $M_1$ is calculated via Eq. 5, while the mass of the secondary component is directly calculated from the estimated mass ratio of the system. 

All the above calculated values are listed in Table~\ref{elements}. 

\begin{table}
\bc
\begin{minipage}[]{100mm}
\caption[]{Estimated absolute elements\label{elements}}\end{minipage}
\setlength{\tabcolsep}{1pt}
\small
 \begin{tabular}{cccc}
  \hline\noalign{\smallskip}
System & J112237 & L1286561 & L2602707 \\
\hline\hline
$a$ & 1.17 & 1.02 & 1.36 \\
$M_{tot}$ & 0.648 & 0.501 & 0.839 \\
$M_1$ & 0.400$\pm$0.010 & 0.112$\pm$0.001 & 0.422$\pm$0.004 \\
$M_2$ & 0.248$\pm$0.030 & 0.389$\pm$0.013 & 0.417$\pm$0.013 \\
$R_1$ & 0.502$\pm$0.004 & 0.244$\pm$0.003 & 0.448$\pm$0.084 \\
$R_2$ & 0.404$\pm$0.007 & 0.459$\pm$0.009 & 0.445$\pm$0.09 \\
$L_1$ & 0.123$\pm$0.001 & 0.005$\pm$0.010 & 0.028$\pm$0.010 \\
$L_2$ & 0.051$\pm$0.003 & 0.015$\pm$0.001 & 0.026$\pm$0.01 \\
$Mbol_1$ & 7.03 & 10.49 & 8.65 \\
$Mbol_2$ & 7.98 & 9.31 & 8.73 \\
$log\rho_1$ & 0.651 & 1.037 & 0.821 \\
$log\rho_2$ & 0.724 & 0.735 & 0.823 \\
$log g_1$ & 4.64 & 4.71 & 4.76 \\
$log g_2$ & 4.62 & 4.7 & 4.76 \\
$log jrel$ & -1.28 & -1.6 & -1.07 \\
  \noalign{\smallskip}\hline
\end{tabular}
\ec
\end{table}

Following the equation of page 131 of \cite{Popper1977} paper, we calculated the orbital angular momentum of the targets
\begin{equation}
J_{rel} = M_1 M_2 (P / M_1 + M_2)^{1/3}.
\end{equation}
where $P$ is in days and $M_i$ are in solar units.

The obtained values $log J_{rel}$ (Tab.6) of our two LINEAR systems are considerably smaller than those of detached systems, which have $log J_{rel}>  +0.08$. 

The orbital angular momentum of J112237 is smaller even than that of contact systems, which have $log J_{rel} >− 0.5.$

The small orbital angular momentum of J112237 implies the existence of a past episode of angular-momentum loss during binary evolution. It also means that J112237 is not a pre-main-sequence object. 

\section{Discussion on the systems}
\label{sect:discussion}

Here we have presented the analysis of filtered CCD light curves of three USPBs. One of these, J112237, is a contact system of spectral type K while the other two are rare detached systems in which the components are non degenerate M dwarf.

\subsection{J112237}
The values of mass ratio found for J112237 indicates that the system is a typical W-subtype contact binary in the Binnendijk  classiﬁcation (\cite{Binnendijk1965};\cite{Binnendijk1970}). The system has a shallow
contact configuration (fill-out 14.8\%).

Contact binaries below the period limit of 0.22 days, i.e. Ultra Short Period Contact Binaries (USPCBs), are expected should be composed of two K-type or later-type components, according to the period-color relation for contact binaries \citep{Zhang2014}. According to the accepted orbital solution, the spectral type of the secondary component of J112237 is estimated to be K5 following \cite{Cox2000}.

It is that a large amount of K-type short-period contact binaries may be W-type systems \citep{Liu2014b} and that the majority of W- type contact binaries show characteristics of being in shallow contact \citep{Zhu2010}.

In recent years the list of known USPCBs, and their study, has been substantially extended, founding that most USPCBs have shallower fill-out factors (f\textit{} \(<\) 20\%) indicating that USPCBs have just evolved to a contact phase. 

According to the thermal relaxation oscillation (TRO) theory (\cite{Lucy1976}: \cite{Lucy1979a}; \cite{Flannery1976}; \cite{Robertson1977}; \cite{Yakut2005} and \cite{Li2008}), via mass transfer between the components, a cycle of contact-semidetached-contact states will be formed. 

The relatively small temperature difference between components of J112237 ($\Delta T = T_h-T_c = 510K$), is accepted for overcontact systems. 

The different light levels at the quadrature were reproduced by a cool spot on the primary component.

\begin{figure*}
\center
\includegraphics[width=12.0cm,height=5.5cm]{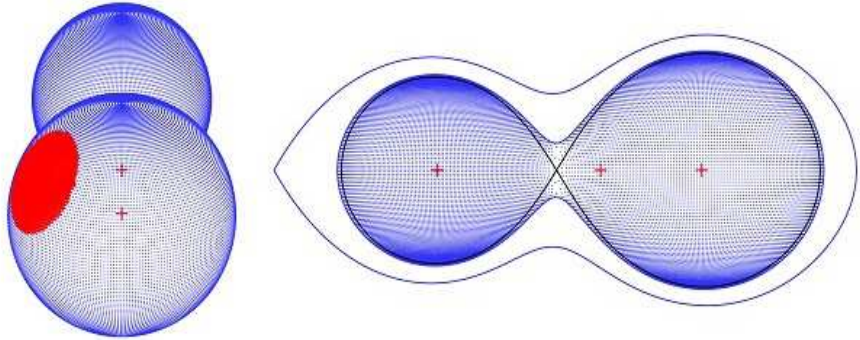}
\caption{Geometric representation of  CSS J112237 at the primary minimum (left) and the configuration of the components of the system in the orbital plane at phase 0.25 (right), according to our solution.}
\label{fig:J112237}
\end{figure*}

A graphic representations and the Roche geometries of J112237 are shown in Figure~\ref{fig:J112237}.

\subsection{L1286561 and L2602707}

The LINEAR systems are two rare M dwarf detached USPB systems with non degenerate components. Among them only other five systems are well or quite well studied to date (see Table~\ref{other}).

\begin{table}
\bc
\begin{minipage}[]{100mm}
\caption[]{List of a quite well studied detached M dwarf systems\label{other}}\end{minipage}
\setlength{\tabcolsep}{1pt}
\small
\footnotemark{}
 \begin{tabular}{cccccccc}
\hline\hline
Name & NSVS 4876238 & BX Tri & BW3 V38 & OGLE BLG ELC-000066 & CSS J171508.5+350658 & L1286561 & L2602707\\
  \hline\noalign{\smallskip}
Period & 0.22184 & 0.19264 & 0.19839 & 0.0984 & 0.17855 & 0.1824 & 0.19973 \\
Sp. Type & K9 & dM & M3 & M4 & M1 & M4.5V & M2 V \\
$T_1$ & 3860 & 3735 & 3500 &   & 3700 & 3120 & 3520 \\
$T_2$ & 3826 & 3106 & 3448 &   & 3600 & 2989 & 3465 \\
\textit{q} & 0.379 & 0.519 & 0.95 & 0.95 & 0.895 & 3.47 & 0.98791 \\
\textit{i} & 68 & 72.5 & 85.51 & 86 & 67.9 & 72.9 & 79.9 \\
$r_1$ & 0.453 & 0.431 & 0.372 &   &   & 0.239 & 0.33 \\
$r_2$ & 0.264 & 0.228 & 0.323 &   &   & 0.449 & 0.328 \\
a & 1.58 & 1.28 & 1.355 &   &   & 1.02 & 1.36 \\
$M_1$ & 0.78 & 0.51 & 0.44 & 0.22 &   & 0.112 & 0.422 \\
$M_2$ & 0.3 & 0.26 & 0.41 & 0.21 &   & 0.389 & 0.417 \\
$R_1$ & 0.72 & 0.55 & 0.51 & 0.26 &   & 0.244 & 0.448 \\
$R_2$ & 0.42 & 0.29 & 0.44 & 0.23 &   & 0.459 & 0.445 \\
$L_1$ & 0.102 & 0.053 & 0.035 &   &   & 0.005 & 0.028 \\
$L_2$ & 0.034 & 0.007 & 0.025 &   &   & 0.015 & 0.026 \\
Reference & 1 & 2 & 3 & 4 & 5 & 6 & 6\\
\hline
\end{tabular}
\ec
{(1)\cite{Kjurkchieva2018}, (2) \cite{Dimitrov2010}, (3) \cite{Maceroni2004} and \cite{Dimitrov2010}, (4)  \cite{Soszynski2015} (5) \cite{Kjurkchieva2016}, (6) This paper.\\}
\footnote{(1)\cite{Kjurkchieva2018}, (2) \cite{Dimitrov2010}, (3) \cite{Maceroni2004} and \cite{Dimitrov2010}, (4)  \cite{Soszynski2015} (5) \cite{Kjurkchieva2016}, (6) This paper.\\}
\end{table}

As discussed in \cite{Becker2011}, the sample of known binary systems composed of two M dwarfs is very small. 

For many years the shortest period known M dwarf binary system in the literature was BW3 V38 (\cite{Maceroni1997}, \cite{Maceroni2004}, composed of two main sequence M3 dwarfs with an orbital period of 0.1984 days
and the similarity between the absolute parameters of this system with those of L2602707 is really surprising. \cite{Dimitrov2010} reported the shortest period M dwarf binary yet characterized, GSC 2314-0530 (BX Tri) with a 0.192636 days period \cite{Soszynski2015} found, among 242 USPBs, OGLE-BLG-ECL-000066 with the orbital period below 0.1d and the two components are M dwarfs in a nearly contact configuration.

Subsequently, \cite{Kjurkchieva2016} report observations of CSS J171508.5+350658 a semidetached system in which both the components consists of M dwarf with period of 0.178d. In the 2108, \cite{Kjurkchieva2018}, during photometric and spectroscopic observations of USPBs found NSVS 4876238 an USPB of detached type in which both components are late dwarf (K9 spectral type) and the temperature difference between the components does not exceed 400K.

For both our systems the difference between the temperature of the components is really poor and, while L2602707 show symmetric light curves, for L1286561, to represent the asymmetries of its light curves, was necessary add a couple of cool spots on the massive component. 

Light asymmetry has been reported commonly for light curves of short-period binaries and may be due to the spot activity on stellar photospheres.  They can be reproduced by surface temperature inhomogeneities (spots). It is reasonable to assume existence of cool spots by analogy with our Sun. An interesting characteristic of L2602707 is that it is a twin binary system in which the mass ratio is near 1 ($q = 0.9879$). These kind of binaries were first noted by \cite{Lucy1979b}.

Statistical studies on the mass ratio distribution of binaries (e.g. \cite{Lucy2006}; \cite{Simon2009}) showed that the frequency of existing twins within the mass ratio 0.98\(-\)1.00 is about 3\% among all binaries, at which F, G, and K spectral type systems dominate.

Among the systems of Table 7, L2602707 is not the only twin binary.

\begin{figure*}
\center
\includegraphics[width=12.0cm,height=7.5cm]{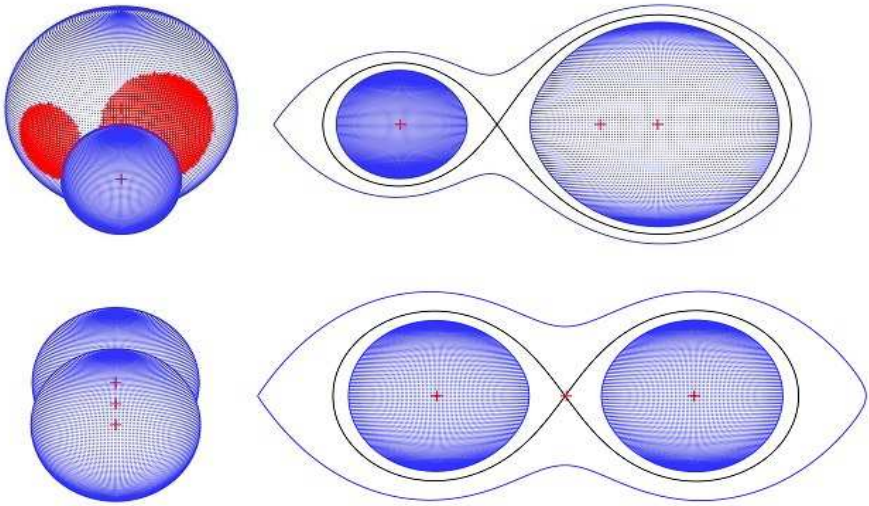}
\caption{As for Figure~\ref{fig:J112237} but for L1286561 (top) and L2602707 (bottom)}
\label{fig:LINEAR}
\end{figure*}

A graphic representation and the Roche geometry of L1286561 and L2602707 are shown in Figure~\ref{fig:LINEAR}.

\begin{figure*}
\center
\includegraphics[width=9.0cm,height=7.0cm]{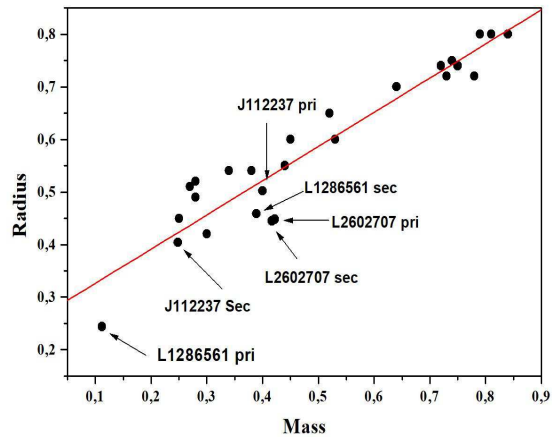}
 \caption{The mass-radius diagram of our three systems with other ten USPBs taken from \cite{Kjurkchieva2018}. The solid line represents the linear fit to the data.}
\label{fig:diagram}
\end{figure*}

\begin{figure*}
\center
\includegraphics[width=9.0cm,height=7.0cm]{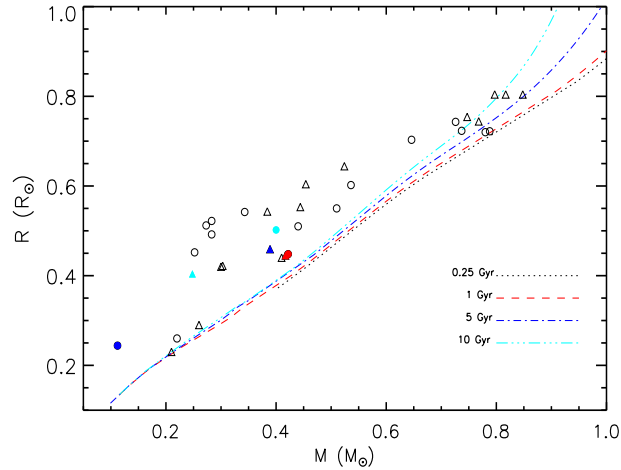}
 \caption{Positions of the components of J112237 (filled cyan symbols), L2602707 (filled red symbols) and L1286561 (filled blue symbols) on the mass-radius diagram compared with ten USPBs from  \cite{Kjurkchieva2018} depicted with open circles for the primary components and open triangles for the secondary components.  Also plotted are theoretical isochrones derived from Dartmouth models \citep{dotter2008} for solar metallicity, with ages 0.25, 1.0, 5.0 and 10 Gyr.}
\label{fig:diagram}
\end{figure*}

The components of all our three systems are shown in the mass-radius diagram (Fig.6) with other ten USPBs systems whose data are taken from \cite{Kjurkchieva2018}.

Additionally, the positions of the components of the three systems are compared  in the mass-radius diagram in Fig. 7 with the USPBs listed by \cite{Kjurkchieva2018} together with theoretical isochrones derived from Dartmouth models \citep{dotter2008} for solar metallicity and ages 0.25, 1.0, 5.0 and 10.0 Gyr.

Our systems follow the general pattern of the USPBs systems.

\begin{acknowledgements}
RM and LFM acknowledge the financial support from the UNAM under DGAPA grant PAPIIT IN 100918.
\end{acknowledgements}


\label{lastpage}

\end{document}